\documentclass[aps,amsmath,twocolumn,amssymb]{revtex4}
\usepackage{graphicx}
\usepackage{bm}
\usepackage{amssymb}
\usepackage{float}
\usepackage{graphicx}
\usepackage{dcolumn}
\usepackage[colorlinks]{hyperref}
\hypersetup{citecolor=blue}
\usepackage{bm}
\usepackage{epstopdf}
\usepackage{amsthm}
\usepackage{xcolor}
\usepackage{float}
\graphicspath{{./figures/}}
\def\phd#1#2#3{{ {Physica D}} {\bf#1},  #2 (#3)}
\def\ph14d#1#2{{ {Physica {\bf14D}}} {#1}, (#2)}

\def\cmp#1#2#3{{{ Commun. Math. Phys.}} {\bf #1}, #2 (#3)}
\def\prl#1#2#3{{ {Phys. Rev. Lett.}} {\bf #1}, #2 (#3)}
\def\pla#1#2#3{{Phys. Lett. A} {\bf #1}, #2 (#3)}

\def\ijbc#1#2#3{{ {Int. J. Bif. Chaos}} {\bf #1}, #2 (#3)}

\def\csf#1#2#3{{{Chaos Soliton \& Fractal}}  {\bf #1}, #2  (#3)}
\begin{document}

\title{A higher-dimensional generalization of the Lozi map: Bifurcations and dynamics}

\author{Shakir Bilal$^1\footnote{This article was written when the author was at University of Delhi. He works from home at the time of submission.}\footnote{email: shakir.bilal@gmail.com}$ and Ramakrishna Ramaswamy$^{2}\footnote{email: r.ramaswamy@gmail.com}$}
\affiliation{$^1$Department of Physics and Astrophysics, University of Delhi, Delhi, 110 007, India\\
$^2$Department of Chemistry, Indian Institute of Technology, 
New Delhi, 110 016, India}

\begin{abstract}
We generalize the two dimensional Lozi map in order to systematically obtain piece--wise 
continuous maps in three and higher dimensions. Similar to higher-dimensional generalizations 
of the related H\'{e}non map, these higher-dimensional Lozi maps support 
hyperchaotic dynamics. We carry out a bifurcation analysis and investigate the dynamics 
through both numerical and analytical means. The analysis is extended to a sequence of approximations that smooth the discontinuity in the Lozi map. 
\end{abstract}
\keywords{Piecewise smooth map, Hyperchaos, Smooth approximation.}
\maketitle

\section{Introduction}\label{sec1}
The behavior of low-dimensional nonlinear iterative maps and flows has been extensively studied and characterized over the past few decades, particularly with reference to the creation of chaotic dynamics \cite{May76, Yorke75, Henon76, Lozi78}.  The various scenarios or routes to chaos in such systems are by now fairly well known \cite{May76, Arnold65, Newhouse78, Rand82}. Similar exploration of the properties of higher dimensional dynamical systems---for instance  the dynamics of attractors  with more than one positive Lyapunov exponent and the bifurcations through which they have been created---has not been studied in as much detail even in relatively simple systems \cite{Lozi78,Sprott06,Albers06,Baier90,Elhadjbook14}. 

Linear and piecewise-linear mappings are among the simplest examples of iterative dynamical systems. The so--called Lozi map \cite{Lozi78} is analogous to the quadratic H\'enon mapping  \cite{Henon76} but has the advantage that more extensive analysis is possible \cite{SCOR11}. The mapping itself is only piecewise continuous, and this introduces some additional features that need to be understood more clearly \cite{bernardobook,Simpson08,Simpson12,Kuznetsovbook}. Indeed, specific bifurcation phenomenon such as border collision bifurcations can only occur in piecewise smooth dynamical systems \cite{Simpson12,bernardobook}. 

Our interest in the present paper is the generalization of the Lozi map to higher dimensions. One motivation is to compare this piecewise continuous system to a similar high-dimensional H\'enon mapping \cite{SBRR13}.  Of the different ways in which this can be done, we choose to extend the map to $d$--dimensions by incorporating time-delay feedback while ensuring that the system remains an endomorphism in the absence of dissipation. The dissipation is introduced at the $k$th step, $k<d$, and this also ensures that the map is a diffeomorphism. The system can therefore have $k$--positive Lyapunov exponents, and we examine the transition to high-dimensional chaos as a function of parameters, characterizing the different bifurcations that can occur. An intermediate ``smooth'' approximation \cite{Aziz01} of the piecewise map is also investigated  vis-a-vis bifurcations for comparison.

In the next Section, the generalized Lozi map is described and a detailed analysis of the  local bifurcations of the elementary fixed points is presented. The emergence of chaotic and hyperchaotic attractors and the global bifurcations that arise are discussed in Section \ref{sec3}. Section \ref{sec4}  is devoted to the analysis of smooth approximations to the map. The paper concludes with a discussion and  summary in Section \ref{sec5}.

\section{The generalized Lozi map}\label{sec2}
The two dimensional Lozi map \cite{Lozi78} is given by
\begin{eqnarray}\label{eq1}
x_{n+1}&=&1-(1-\nu)y_n-a|x_n|\nonumber\\
y_{n+1}&=&x_n\;.
\end{eqnarray}
This map is a modification of the quadratic H\'enon mapping, with the parameters $\nu$ and 
$a$ tuning the dissipation and nonlinearity respectively. Since the map is piecewise linear, it lends itself
to extensive analysis, some of which has been recently summarized \cite{Elhadjbook14}.

\begin{figure}[ht]
\centering
{\scalebox{0.4}{\includegraphics{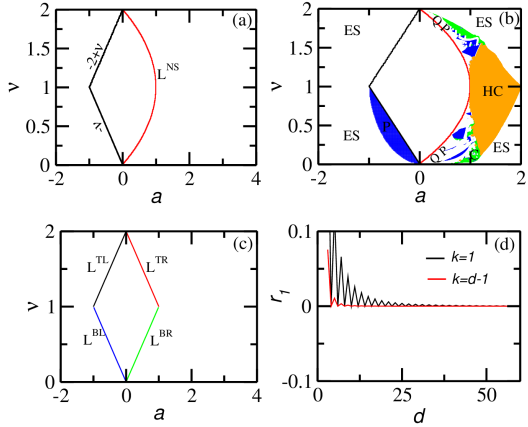}}}
\caption{(a) Region of period--1 dynamics for $d$=3, $k$=2; see Eq.\eqref{eq7}. (b)The organization of $a$--$\nu$ plane into unbounded dynamics (ES) and bounded dynamics: hyperchaotic (HC,orange), chaotic (C, green), periodic (P,blue). White regions sandwiched between period one dynamics and hyperchaotic/chaotic/periodic corresponds to quasiperiodic dynamics (QP). The blue regions indicate that supercritical bifurcations are possible on either side of the period one boundaries (see Fig.~\ref{Fig2}).
(c) The region of period one dynamics converges asymptotically (i.e. for $d\rightarrow \infty$) and is contained within the curves $L^{TL},L^{TR},L^{BL},L^{BR}$. (d) The convergence to the curves TR,BL,BR is captured by the distance  $r_1$ of the largest root from the unit circle as a function of dimension $d$ on these curves. $L^{NS}$ indicates the 
Neimark--Sacker boundary, and the other labels $L^{TR}$, $L^{TL}$, $L^{BR}$ and $L^{BL}$ are short for top left,right, and bottom left,right.}
\label{Fig1}
\end{figure}
Rewriting the above as a difference delay equation,  one has 
\begin{eqnarray}\label{eq01}
x_{n}&=&1-a|x_{n-1}|-(1-\nu)x_{n-2},
\end{eqnarray}
which suggests a natural generalization to higher dimensions, 
\begin{eqnarray}\label{eq2}
x_{n}&=&1-a|x_{n-k}|-(1-\nu)x_{n-d}.
\end{eqnarray}
Here $d$ and $k$ are integers such that $k<d$, and we take $0\leq\nu\leq 2$. The mapping is conservative when $\nu$  is 0 or 2,  and is 
dissipative otherwise.  For $\nu=1$ the map reduces to a $k$--dimensional endomorphism,  while for $\nu\neq 1$ the map is a 
$d$--dimensional diffeomorphism.  In the next subsection we analyze the implications of different choices of $d$  and $k$ for this map.

\subsection{The base maps and $q$--degeneracy}\label{sec2p1}
The integers $d$ and $k$ are either co-prime or share a common factor $q$. When $k$ and $d$ have a common factor $q$, 
it is easy to see that all the eigenvalues of the Jacobian are $q$--fold degenerate, and this leads to a $q$--fold degeneracy in the
Lyapunov exponents. It therefore suffices to examine the case of $d, k$ co--prime since these form the {\em base} for all other values of $k<d$,
and it suffices to consider only base-maps as can be seen by the following argument. For the $n=qm$th iterate,
the substitution $x_{q\cdot}$ by $\xi_\cdot$ gives
\begin{eqnarray}\label{eq3}
x_{qm}&=&1-a|x_{qm-qk'}|-(1-\nu)x_{qm-qd'}\nonumber\\
\downarrow\nonumber\\
\xi_{m}&=&1-a|\xi_{m-k'}|-(1-\nu)\xi_{m-d'}.
\end{eqnarray}
The maps Eq.~\eqref{eq2} and Eq.~\eqref{eq3} differ in that there are $q$ hidden variables 
within each $\xi_m$. Thus the Jacobian can be separated into $q$ identical blocks,
giving rise to a $q$--degenerate $(d',k')$ map.
 
\subsection{Fixed points: Stability}\label{sec2p2}
The delay map, Eq.~\eqref{eq2} can be rewritten as a $d$-dimensional iteration, 
\begin{eqnarray}\label{eq02}
x^1_{n+1}&=&1-a|x^k_{n}|-(1-\nu)x^d_{n}\nonumber\\
x^2_{n+1}&=&x^1_{n}\nonumber\\
\vdots\nonumber\\
x^{d}_{n+1}&=&x^{d-1}_n
\end{eqnarray}
The fixed points of the mapping are those for which $\{x^1_{n+1},\ldots,x^d_{n+1}\} = \{x^1_{n},\ldots, x^d_{n}\}$. Solving, we find 
\begin{eqnarray}\label{eq4}
x^{1*}&=&x^{2*}=\dots=x^{d*}=x_{\pm}\nonumber\\
\mathrm{with~~}x_{\pm}&=&\frac{1}{2-\nu\pm a}.
\end{eqnarray}
of these two fixed points, $x_-$ is always unstable. The matrix elements of the Jacobian of the map in Eq.~\eqref{eq2} are
 given by 
\begin{eqnarray}
\label{eq5}
\left[ \mathbf{J}\right]_{ij} = \begin{cases} 1 & j=i-1\;, \;\;d\;\geq\;i\;\geq\; 2 \\
\pm a &  i=1\;,\; j=k \\
-(1-\nu) & i=1\;,\; j=d \\
0 & \mathrm{otherwise} \end{cases}\;,
\end{eqnarray}\noindent
and it is straightforward to obtain the stability conditions on the fixed point $x_{+}$ (for arbitrary $d$ and $k$) 
from the characteristic polynomial $P(\lambda)$,
\begin{eqnarray}\label{eq6}
P_{1,2}(\lambda)&=&\lambda^d \;\pm\; a\lambda^{d-k}+(1-\nu).
\end{eqnarray}\noindent 
If the number of real roots of the polynomials $P_{1,2}$ with real part greater than +1 (or smaller than -1)
is $\sigma^+_{1,2}$ ($\sigma^-_{1,2}$), then according to Feigin's classification of border collision bifurcations in 
piecewise smooth maps \cite{bernardobook}, a fold bifurcation occurs when $\sigma^+_1\;+\;\sigma^+_2$
is odd. If $\sigma^-_1\;+\;\sigma^-_2$ is odd, on the other hand, a flip bifurcation occurs. 

The characteristic of the Neimark--Sacker bifurcation in smooth dynamical systems is that a pair of complex 
eigenvalues cross the unit circle. This theory has been extended to piecewise smooth maps only recently \cite{Simpson08}
and although some of the features are common, there are major differences \cite{Simpson08}. 
We find from numerical estimation of eigenvalues of $P_{1,2}$ for the base maps that a Neimark--Sacker
 bifurcation occurs via a border collision  if an odd number of pairs of complex eigenvalues cross the unit circle. 
 In particular, for the base map with $d=3, k=2$ we find that period one region is 
 bounded by the curves as shown in Fig.~\ref{Fig1}(a):
 \begin{eqnarray}\label{eq7}
L^{\mathrm{fold}}(d)\;:\;a&=&-2+\nu\nonumber\\
L^{\mathrm{NS}}(d=3)(\mathrm{supercritical})\;:\;a&=&\nu(2-\nu)\;.\nonumber\\
L^{\mathrm{flip}}\;:\;a&=&-\nu\;.\nonumber
 \end{eqnarray}

If either of the polynomials produce their largest roots with absolute values less than one, 
the fixed point $x_+$ is stable and contributes to the period--one region in the $a$--$\nu$ 
parameter space, unless it hits the border $x=0$. Unlike the smooth case (i.e. the H\'{e}non map) 
studied in \cite{SBRR13} the bifurcations in the Lozi map can show supercritical bifurcations 
on either side of the period one boundaries (of course limited upto the saddle node curve), as shown in Fig. \ref{Fig1}(b). 
Since these bifurcation clearly show that orbit of the period one hits the boundary $x=0$ at 
such bifurcation, we attribute this distinct feature of the generalized version of the Lozi map 
\eqref{eq2} to border collision bifurcations of the map.
\begin{figure}[ht]
\centering
{\scalebox{0.4}{\includegraphics{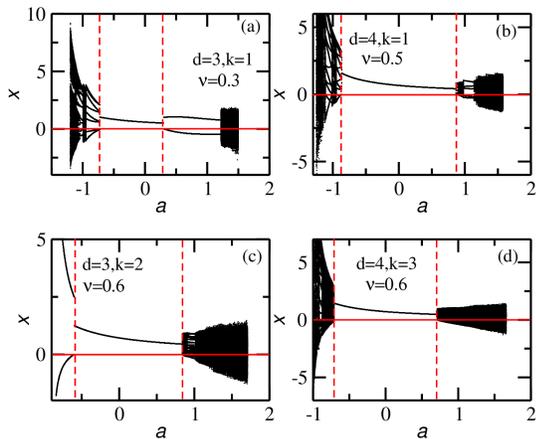}}}
\caption{Bifurcation diagrams as a function of the nonlinearity $a$ for different base maps 
(a) $d$=3, $k$=1, $\nu$ = 0.3 (b) $d$=4, $k$=1, $\nu$=0.5, (c) $d$=3, $k$=2, $\nu$=0.6, (d) $d$=4, $k$=3, $\nu$=0.6. 
The boundaries of stable period-1 dynamics are indicated by dashed vertical lines; these are also the border-collision bifurcation points.}
\label{Fig2}
\end{figure}
\subsection{Bifurcations diagrams}\label{sec2p3}
\begin{figure}[ht]
\centering
{\scalebox{0.4}{\includegraphics{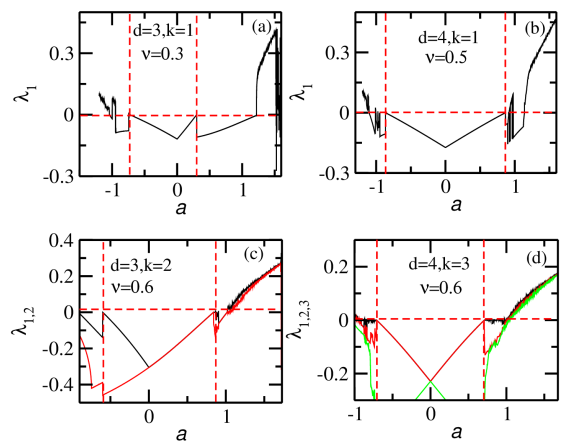}}}
\caption{k- largest Lyapunov exponents for the systems corresponding to the bifurcation diagrams shown in Fig.~\ref{Fig2} . The different base maps are
(a) $d$=3, $k$=1, $\nu$=0.3, (b) $d$=4, $k$=1, $\nu$=0.5, (c) $d$=3, $k$=2, $\nu$=0.6,  and (d) $d$=4, $k$=3, $\nu$=0.6. 
The boundaries of the stable period-1 dynamics are indicated by dashed vertical lines; the left boundary is a border collision bifurcation point.}
\label{Fig3}
\end{figure}

Bifurcation diagram in the two dimensional $a$--$\nu$ parameter space for $d=3,k=2$ is shown in Fig.~\ref{Fig1}(b). A typical feature of $a$--$\nu$ parameter space is the existence of hyperchaotic (HC), chaotic (C), quasiperiodic (QP), and periodic regions. These features are shared by other members (with different $d$ and $k$ values) of this generalized Lozi map. Representative bifurcation diagrams as a function of $a$ for different $(d, k)$ combinations are shown in Fig.~\ref{Fig2} and the corresponding orbital characteristics are shown via Lyapunov exponents in Fig.~\ref{Fig3}. In each case $\nu$ is different but fixed. An interesting feature of these bifurcation diagrams is that for for $0<\nu<1$ some base maps exhibit bounded non--trivial dynamics as the parameter $a$ is decreased to the left of the period one boundary. Typically these were found to be a flip bifurcation below $a<-\nu$ when $d=3,k=2$ and supercritical Neimark--Sacker type bifurcations for $d=4,k=3$. Such feature are typically of these maps even across different combination of dimensionality parameters ($d$,$k$). We should mention that such phenomenon was not found for a similarly generalized H\'{e}non map \cite{SBRR13}, and appears to be a result of border collision bifurcations. It is important to note that the theory of bifurcations in smooth dynamical systems does not explain these features \cite{Kuznetsovbook}.
 
\section{High dimensional dynamics}\label{sec3}

In this Section we examine the bifurcations starting from the 
period-1 fixed point as a function of nonlinearity parameter for different embedding dimensions $d$ 
and the endomorphism dimension $k$.
\subsection{Bounded dynamics}\label{sec3p1} 
In the limit $d\rightarrow \infty$ period--1 motion converges to a region shown in Fig.~\ref{Fig1}(c): the boundaries are the following curves, 
\begin{eqnarray}
\label{eq8}
L^{TL}\:\: :\;a&=&-2+\nu\nonumber\\
L^{TR}\:\: :\;a&=&2-\nu\nonumber\\
L^{BL}\:\: :\;a&=&-\nu\\
L^{BR}\:\: :\;a&=&\nu,\nonumber
\end{eqnarray}
in the $a-\nu$ parameter space. These curves can be understood from the properties of the characteristic polynomials Eq.~\eqref{eq6} in the limit of $d\rightarrow \infty$. The distance $r_1$ between the leading root of the characteristic polynomial Eq.~\eqref{eq6} and the unit circle on the curve $L^{TL}$, in the extreme case of $k=1$ and $k=d-1$, are shown in Fig.~\ref{Fig1}(d): $r_1$ approaches zero as the dimension $d$ is increased. Similar behavior of $r_1$ is also observed on $L^{BR},L^{BL}$, $T^{LR}$ and for $1<k<d-1$.

\subsection{Hyperchaos}\label{sec3p2}

\begin{figure}[ht]
\centering
{\scalebox{0.4}{\includegraphics{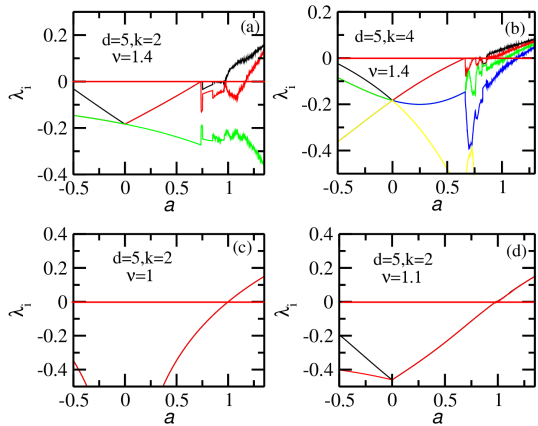}}}
\caption{The Lyapunov exponent spectrum ($\lambda_i,i=1,...,k+1$) as a function of nonlinearity parameter for $d=5$. (a)--(b) demonstrate that for a given dimension $d$ the number of positive Lyapunov exponents are governed by the parameter $k$. We keep $\nu=1.4$ and plot all the Lyapunov exponents for $k=2$ in (a) and $k=4$ in (b). (c)--(d) demonstrate the breaking of degeneracy in the Lyapunov spectrum observed in the endomorphism $\nu=1$ by making the map diffeomorphic $\nu=1.1$ (see text for explanation).}
\label{Fig4}
\end{figure}

The map Eq.~\eqref{eq2} exhibits at most $k$ positive Lyapunov exponents as nonlinearity parameter $a$ is varied: this is due to the fact that the 
nonlinearity in the map occurs at the $k^{th}$ previous iteration step. The 
stretching and folding  that is responsible for introducing sensitivity to initial conditions in the map \cite{Ottbook}, 
occurs in $k$ directions, and this results in the maximum $k$ number of possible positive Lyapunov exponents; see Figs.~\ref{Fig4}(a)--(b). Additionally, for $\nu=1$, the map is a $k$--dimensional endomorphism (and is not invertible) 
with $k$--fold degenerate Lyapunov exponents, i.e. all of these LEs are identical and become positive at the same value of the nonlinearity parameter $a$ as can be seen in Fig.~\ref{Fig4}(c). Embedding the $k$--dimensional endomorphism in a $d$--dimensional space does not change this behavior. However when the map is made diffeomorphic by enabling the contraction/dissipation parameter  ($\nu\neq 1 \;\&\; 0\leq\nu \leq 2$) the degeneracy in the Lyapunov exponents is lifted, although the maximum possible number of positive Lyapunov exponents is still $k$ as can be seen in Fig.~\ref{Fig4}(d). 
 
 Route to chaos is observed via the quasiperiodic and also via finite period--doubling route, as seen in the Lyapunov spectra Fig.~\ref{Fig3} 
 and Fig.~\ref{Fig4}. The period doubling cascade terminates after a few doublings, leading to chaos.
 On the other hand chaos and hyperchaos transition is smooth, since the first $k$--largest Lyapunov exponents behave smoothly as they hierarchically become positive at different values of the nonlinear parameter $a$. This typically means that the map Eq.~\eqref{eq2} can be written as a hierarchy of chaotically driven maps at subsequent transitions to higher chaos \cite{Harr00}, this is similar to the chaos hyperchaos transition in the generalized H\'{e}non map \cite{SBRR13}.

\section{Smooth approximations}\label{sec4}
In this section we analyse a smooth approximation of the generalized Lozi map \eqref{eq2}.
\begin{figure}[ht]
\centering
{\scalebox{0.4}{\includegraphics{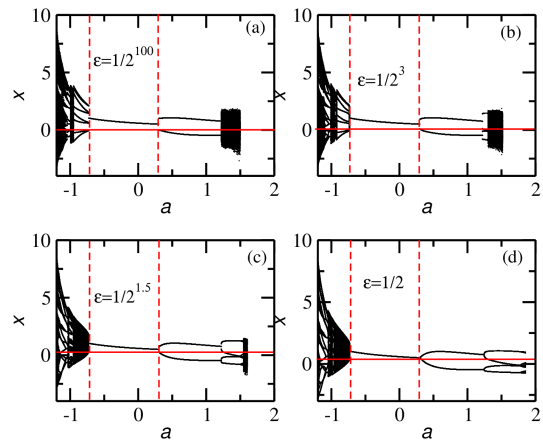}}}
\caption{The bifurcation diagram for $d=3,k=1$ for different values of $\epsilon$. As $\epsilon$ is increased from (a)--(d) characteristics of smooth bifurcations emerge in the form of period doubling.}
\label{Fig5}
\end{figure}
  We replace the modulus function $|\cdot|$ in Eq.~\eqref{eq2} with a smooth function $S_\epsilon(\cdot)$:
\begin{eqnarray}\label{eqsmooth}
x_{n}&=&1-aS_\epsilon(x_{n-k})-(1-\nu)x_{n-d}\\
S_\epsilon(x_{n-k})&=&\begin{cases} x_{n-k}^2/2\epsilon + \epsilon /2 & if\;\; |x_{n-k}|\leq \epsilon \\
 |x_{n-k}| &  if \;\; |x_{n-k}|\geq\epsilon
\end{cases}\;,\nonumber
\end{eqnarray}

where $0<\epsilon<1$. The function $S_\epsilon(x_{n-k})$ extends the smooth approximation applied to the two dimensional Lozi map \cite{Aziz01} to our high dimensional generalization of the Lozi map and removes the discontinuity in the slope at $x_{n-k}=0$. 

The fixed points of the new map in Eq.~\eqref{eqsmooth} are given by:
\begin{eqnarray}\label{fpsmooth}
x_{\pm}&=&
\begin{cases}
\frac{1}{2-\nu\pm\;a} & if\;\; |x|>\epsilon\\
 \frac{\epsilon}{a}\Big(-(2-\nu)\\
 \pm\sqrt{(2-\nu)^2+\frac{2a}{\epsilon}-a^2}\Big)& if\;\; |x|\leq \epsilon
\end{cases}
\end{eqnarray}
the first set of these fixed points are similar to those of Lozi map (see Eq.~\eqref{eq4}) and lose stability by colliding with one of the borders located at $\pm\epsilon$ as nonlinearity parameter $a$ is varied. The new orbits that appear following this border collision bifurcation depend on the delay parameters $d$ and $k$, although the maximum number of positive Lyapunov exponents is still limited to $k$. Assuming fixed values of the dissipation parameter and $\epsilon$, the variation in nonlinearity parameter can take iterations of the map also inside the region $|x|<\epsilon$ then subsequent bifurcations are no longer only due to border collisions: borders at $\pm\epsilon$ have well defined first derivatives and once an orbit enters the region $|x|<\epsilon$ the dynamics is also governed by the smooth approximation.

The case of $d=2$ and $k=1$ was illustrated in \cite{Aziz01}, where it was observed that the period-doubling route to chaos is achieved for finite values of $\epsilon$ (note that period doubling route to chaos is absent for $\epsilon$=0).

\begin{figure}[ht]
\centering
{\scalebox{0.4}{\includegraphics{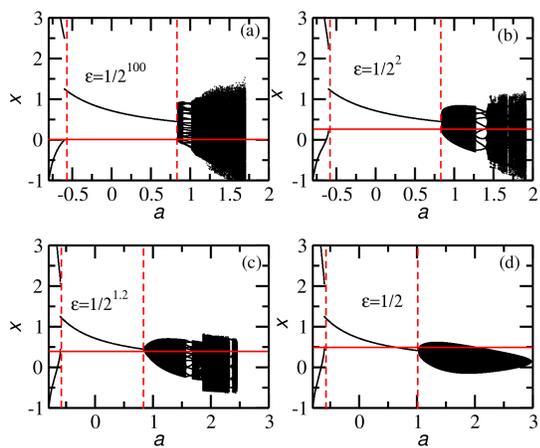}}}
\caption{The bifurcation diagrams for $d=3,k=2$ for different values of $\epsilon$. As $\epsilon$ is increased from (a)--(d) characteristics of smooth bifurcations emerge in the form of period doubling. The figures (d)--(a) also indicate emergence of border collision bifurcation beginning from smooth period doubling bifurcations in (d).}
\label{Fig6}
\end{figure}

For $d>2$ bifurcation scenarios easily understood by writing the smoothness parameter as $\epsilon=1/2^{n_\epsilon}, n_\epsilon\ge0$. 
For the case of $d$=3, $k$=1 the loss of stability of the fixed point as nonlinearity parameter is varied is dependent on $n_\epsilon$: border collision bifurcations gives way to period doubling bifurcations as $n_\epsilon$ decreases (Fig.~\ref{Fig5}(a)--(d)). Similar observations are made when a Neimark--Sacker type bifurcation is involved in the Lozi map with $d=3,k=2$, where border collision bifurcations (Fig.~\ref{Fig6}(a) ) give way to smooth bifurcations Fig.~\ref{Fig6}(d).
In the H\'{e}non maps the chaotic dynamics follows directly after the Neimark--Sacker bifurcation via a crisis--like transition \cite{SBRR13}. 
Thus the effect of $\epsilon >$ 0 is to introduce bifurcations which are a mix of border collisions and smooth bifurcations. 
\section{Discussion and Summary}
\label{sec4p5}
\label{sec5}

In this paper we have introduced a generalized time--delayed Lozi map with nonlinear feedback from 
$k$ earlier steps and linear feedback from $d$ earlier steps. This simple feedback process governs the dimensionality of the maps. The parameter $k$ ($<d$) determines the number of positive Lyapunov exponents. Further more, the family of maps formed as a result of different combinations of dimensionality parameters are classified into base 
maps when $d$ and $k$ are co--prime. All other maps reducible to these base maps exhibit a $q$-fold degenerate Lyapunov spectrum when $d$ and $k$ share a common factor $q$. Bifurcation analysis was performed in a limited region of the parameter space. In particular, fixed point dynamics loses stability through the fold, flip and the Neimark Sacker bifurcations via border collisions. Analytic forms were determined for these boundaries, and the flip 
and NS bifurcation curves were found to depend on the dimension $d$. With increasing dimension, the 
region of period one dynamics was found to converge in the parameter plane.  

The dynamics evolves abruptly from regular to chaos due to piece-wise nature of the map. Subsequent transitions from chaos to hyperchaos, however, are smooth as indicated by Lyapunov spectrum: the dimension of the attractor changes smoothly if there are no abrupt transitions in Lyapunov spectrum \cite{Harr00, SBRR13}.

A smooth approximation of the map enabled the analysis of the bifurcations vis-a-vis further comparing some of the bifurcations to the generalized the H\'{e}non map. It showed that some of the bifurcations observed  persist on both the piecewise Lozi and H\'{e}non map. Further exploration of a more general unified mapping is a project for future work. Another  possible project for a future work is the analysis of conservative limit in these class of maps: orbits in the conservative limit are only possible when $d=2k$,therefore in the conservative limit hyperchaotic orbits are indeed possible with $k-$positive Lyapunov exponents. The exploration of the conservative limit of this map could be a task for future work.
\section*{Acknowledgement}
 SB was supported by the UGC (Govt of India), through Dr. D. S. Kothari postdoctoral fellowship at the time of writing this manuscript and RR is supported by the DST (Govt of India) through the JC Bose fellowship. 

\end{document}